\documentstyle[12pt,twoside,psfig]{article}
\begin{document}

\begin{center}
{\Large\bf Acceleration of the universe with a simple \\trigonometric potential 
}
\\[15mm]
Narayan Banerjee \footnote{E-mail: narayan@juphys.ernet.in}~~~~ 

Sudipta Das \footnote{E-mail:dassudiptadas@rediffmail.com}\\
{\em Relativity and Cosmology Research Centre,\\Department of Physics, Jadavpur University,\\ Calcutta - 700 032, \\India}\\[15mm]
\end{center}

\vspace{0.5cm}
\vspace{0.5cm}

\pagestyle{myheadings}
\newcommand{\be}{\begin{equation}}
\newcommand{\ee}{\end{equation}}
\newcommand{\bea}{\begin{eqnarray}}
\newcommand{\eea}{\end{eqnarray}}

\begin{abstract}
In this paper, we investigate the quintessence model with a  
minimally coupled scalar field in the context of recent
supernovae observations. By 
choosing a particular form of the deceleration parameter $q$ , which gives an 
early deceleration and late time acceleration for dust dominated model, we show
that this sign flip in $q$ can be obtained by a simple trigonometric
potential. The early matter dominated model expands with $q = 1/2$ as desired 
and  enters a negative $q$ phase quite late during the evolution.
\end{abstract}

\section{INTRODUCTION}
Over the last few years, there are growing evidences in favour 
of the scenario that the universe at present is expanding with an 
acceleration. The supernovae project \cite{perl} and also the 
Maxima \cite{netter} and 
Boomerang \cite{balbi} data on cosmic microwave background (CMB) strongly 
suggest this acceleration. The very recent WMAP data \cite{dns} also seem 
to confirm this. This result is indeed counter-intuitive, as gravity holds 
matter together and it might be expected that in the absence of any exotic 
field, the universe should be decelerating. As this acceleration could be 
brought about by an effective negative pressure, the first choice of candidate 
for this  `dark energy' had been the `cosmological constant' or a time
varying cosmological parameter $\Lambda(t)$. Due to well-known 
reasons, $\Lambda$ has fallen from grace (for an excellent uptodate review, 
see \cite{vs},\cite{peddy}). A scalar field with a positive definite 
potential can indeed give rise to an effective negative pressure if the 
potential term dominates over the kinetic term. The pressure to density 
ratio for the scalar field, as required by the supernovae observations, 
is given as $w_{\phi} < -\frac{2}{3}$ (See ref \cite{macorra} and references therein). 
This source of energy is called 
the quintessence matter (Q-matter).
 In this context, non-minimally coupled scalar fields had been 
investigated thoroughly to check if they could drive an accelerated expansion 
\cite{berto}. Brans-Dicke's scalar field appears to generate sufficient acceleration 
in the matter era, but it has its problems in the earlier evolution \cite{nb}.
A viscous fluid along with a Q-matter could also be a useful candidate and 
this appears to solve the coincidence problem also \cite{chimento}. This
coincidence problem, i.e, why the Q-matter dominates only recently,
was tackled  in the so called tracker 
solutions \cite{caldwell} where the scalar field energy density runs parallel to the matter energy density from below through the evolution and gets to dominate only during later stages. Very recently Chaplygin gas, which has a nonlinear 
contribution of the energy density to the dynamics of the model, has also been invoked \cite{aas}. 
Most of these models do exhibit an accelerated expansion in the 
matter-dominated regime.\\
\par  It deserves mention that the same model should have a deceleration
in the early phase of matter era 
in order to provide a perfect ambience for structure formation.
Furthermore, the accelerated phase is perhaps only a very recent one. There
are observational evidences too that beyond a certain value of the 
redshift ($z \sim 1.7$), 
our universe had been going through a decelerated expansion \cite{reiss}. 
This indication is indeed reassuring, as the formation of structure in the 
universe is better supported by a decelerating model. This is because 
local inhomogeneities will grow and become stable from the seeds of density 
fluctuation only if the force field is attractive.\\

\par Amendola \cite{amen} has argued that all the required structure 
formation and other relevant observations regarding the supernovae could
well be explained even if the alleged acceleration of universe started 
quite a long time back, even beyond $z = 5$. However, this work also shows 
that the model requires both an accelerated and a decelerated phase of
expansion. But a more recent work by Padmanabhan and Roychowdhury 
\cite{ptr} shows a striking result. It indicates that if we take the 
complete data set, i.e, acceleration upto a certain $z$ and deceleration
beyond that (i.e, for higher $z$), then only this conclusion of the 
change of signature of 
the deceleration parameter holds. On the other hand, the individual 
data sets of the high and low redshift supernovae may well be consistent 
with a decelerating universe without any `dark energy'.\\
    
       So indeed we are in need of some form of fields which governs the
dynamics in such a way that the deceleration parameter becomes negative 
well into the matter era. One such Q-matter had been given by Sen and
Sethi \cite{ss} where they include a potential which is a `double exponential'
of the scalar field. They obtained a scale factor which is a sine hyperbolic 
function of time in the matter dominated regime. The deceleration parameter 
$(q)$ indeed has a sign flip and with a little fine-tuning, the scale factor 
can grow during the early stages as $ t^{2/3}$ which is indeed the usual 
solution for the Einstein equations for a flat FRW spacetime for 
pressureless dust.\\

             In the present work, we adopt the following strategy. We choose a 
form of $q$ as a function of the scale factor  $a$ so that it has the desired 
property of a signature flip. Then with this input, the scalar field and the 
required potential are found out. It turns out that a fairly simple 
trigonometric
potential does the needful. The origin of the scalar potential, however, cannot 
be indicated.
Surely this is not the ideal way to find out the dynamics of the universe, as
here the dynamics is assumed and then the fields are found out without any 
reference to the origin of the field. But  in the
absence of more rigorous ways, this kind of investigations collectively might 
finally indicate towards the path where one really has to search. This
{\it`reverse'} way of investigations had earlier been used extensively by 
Ellis and Madsen \cite{em} for finding out the potential driving inflation,
i.e, an accelerated phase of the universe at a very early stage of its 
evolution.
\noindent
\section{Results}
\par For a spatially flat Robertson-Walker spacetime
\be
ds^2  = dt^2 - a^{2}(t) [dr^2 + r^2 d\Omega^2],
\ee 
the deceleration parameter $q$ is given by
\be
q = - \frac{ \ddot{a} a}{ \dot{a}^{2}}
\ee
where $a$ is the scale factor of the universe and is a function of 
the cosmic time `$t$' alone.\\ In order to get a model consistent with
observations, one needs an expanding universe, i.e, a positive Hubble
parameter $ H = \frac{\dot{a}}{a} $ throughout the evolution, but a 
deceleration parameter $q$, unlike being a positive constant throughout 
the matter era at $ q = 1/2 $ as believed until the recent observations,
should be a function of the scale factor ( or that of $t$ ). Furthermore, this 
functional dependence should be such that $q$ undergoes a transition from its 
positive phase to a negative one in the matter dominated period itself. It is
thus imperative that the scale factor cannot have a simple power-law behaviour. 
If $ a \sim t^{n}$, the universe will have an accelerated or a decelerated 
expansion for $ n > 1 $ or $ n < 1 $ respectively throughout the period.\\

  In the quest for a varying $q$ consistent with observations, in the same line 
  as that floated by Ellis and Madsen, we propose the relation
\be
q = - \frac{ \ddot{a}/{a}}{ \dot{a}^2/ a^2} = - 1 - \frac{ pa^p}{1+a^p},
\ee
where $p$ is a constant. It is found that for a certain range of negative 
values of $p$, this works remarkably well. 

 The equation (3) integrates to yield
\be
H = \frac{ \dot{a}}{a} = A(1+a^p)
\ee
where $A$ is an arbitrary constant of integration. $A$ is taken to be 
positive, which ensures the positivity of the Hubble parameter (the expansion
of the universe is never denied!) irrespective of the signature or value of the 
constant $p$.

\begin{figure}[!h]
\mbox{\psfig{figure=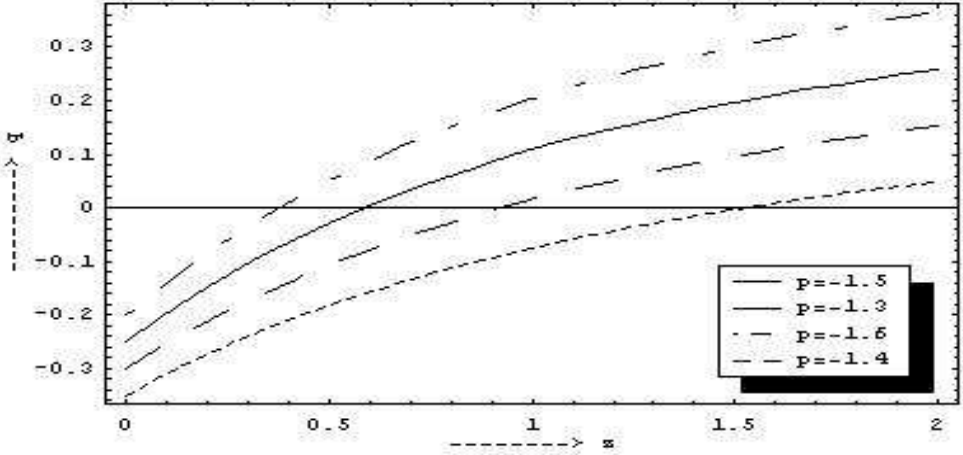,height=2.5in,width=5.3in}}
\caption{Plot of $q$ vs. $z$ for spatially flat dust dominated R-W model.}
\label{first_fig}
\end{figure}

~It is found that for values of $p$ between -2 and -1, the model shows
exactly the behaviour which is desired (as shown in Figure \ref{first_fig}). 

\par In what follows, we work out the problem completely
for $p=-3/2$, for which 
the model works with a non minimally coupled scalar field with a potential
expressed as a simple trigonometric function of the scalar field.

 As the interest is in a matter dominated universe, the fluid is taken in 
the form of a pressureless dust.
The Einstein equations for the space-time given by equation (1) are,
\bea
3\frac{ \dot{a}^2}{a^2} = \rho + \frac{1}{2} \dot{\phi}^2 + V(\phi),\\
2\frac{\ddot{a}}{a} + \frac{ \dot{a}^2}{a^2} = - \frac{1}{2} \dot{\phi}^2 + V(\phi)
\eea
where $\rho$ is the density of matter, $\phi$ is the scalar field and
$V(\phi)$ is a scalar potential.\\
The wave equation for the scalar field is
\be
\ddot{\phi} + 3\frac{\dot{a}}{a}\dot{\phi} + V'(\phi) = 0
\ee
In all equations an overhead dot implies differentiation w.r.t. time and 
a prime is that w.r.t. the scalar field $\phi$. The matter conservation 
equation, which can in fact be obtained from these three equations in view 
of the Bianchi identity, yields
\be
\rho = \frac{ \rho_{0}}{a^3},
\ee
$\rho_{0}$ being a constant. So we have three equations to solve for four unknowns.
\par We assume the deceleration parameter as given in equation (3), which can be 
integrated twice to give  $H = \dot{a}/a$ as in equation (4) and the scale 
factor as
\be
a = [e^{-Apt}-1]^{-\frac{1}{p}}
\ee
\\ Now, the system of equations (5), (6) and (7) is closed with the assumption of equation (3) or equivalently equation (4). So in order to solve the system completely, the parameter $`p$' should have a fixed value. We choose $p=-\frac{3}{2}$, as it yields $q = 0.5$ for a very low value of $\frac{a}{a_0}$, where $a_0$ is the present value of the scale factor. The physical motivation for choosing this value of $q$ for early matter dominated epoch is that for a spatially flat FRW model with $p = 0$ without any Q-matter indeed has $q = 0.5$ and that the transition from radiation to matter dominated epoch for this value of $q$ is well studied \cite{coles}.

 ~ With $p=-\frac{3}{2}$, equations (5) and (6) are used to eliminate  $V(\phi)$, and 
$\dot{\phi}$ can be calculated to be 
\be
\dot{\phi} = \sqrt{3} A a^{\frac {-3}{4}} = \frac{\sqrt{3} A}{[e^{\frac{3At}{2}}-1]^{\frac 1 2}}
\ee
\\
 The scalar field is found out by integrating equation (10) as,
\be
\phi = \frac {4}{\sqrt{3}}  tan^{-1}(e^{+3At/2}-1)^{\frac 1 2},
\ee

The potential $V(\phi)$ can also be calculated from equations (5) and (6)
first as a function of time and by the use of equation (11) as a function 
of $\phi$ as,
\be
V(\phi)  = \frac{9A^2}{2} cot^2(\frac{\sqrt{3}\phi}{4}) + 3A^2
\ee

Now one has the complete set of the solutions, $a = a(t)$, $\phi = \phi(t)$,
$\rho = \rho(t)$ and $V = V(\phi)$  for $p = -3/2$.  The solutions, when
plugged in the field equations, namely (5), (6) and (7), satisfy all of
them provided
\be
\rho_{0} = 3A^2.
\ee
\par From equations (5) and (6), we note that the contribution from the quintessence field $\phi$ towards the density and effective pressure are given as 
\be
\rho_{\phi} = {\frac 1 2}{\dot{\phi}^2} + V(\phi),
\ee
and
\be p_{\phi} = {\frac 1 2}{\dot{\phi}^2} - V(\phi)
\ee
respectively. From these, one can write down the expressions for the dimensionless density parameters $\Omega_{m} = \frac{\rho_{m}}{3H^2}$ and $\Omega_{\phi} =
\frac{\rho_{\phi}}{3H^2}$ respectively for the visible matter and the $Q$-matter.\\

 \par
\begin{figure}[!h]
\mbox{\psfig{figure=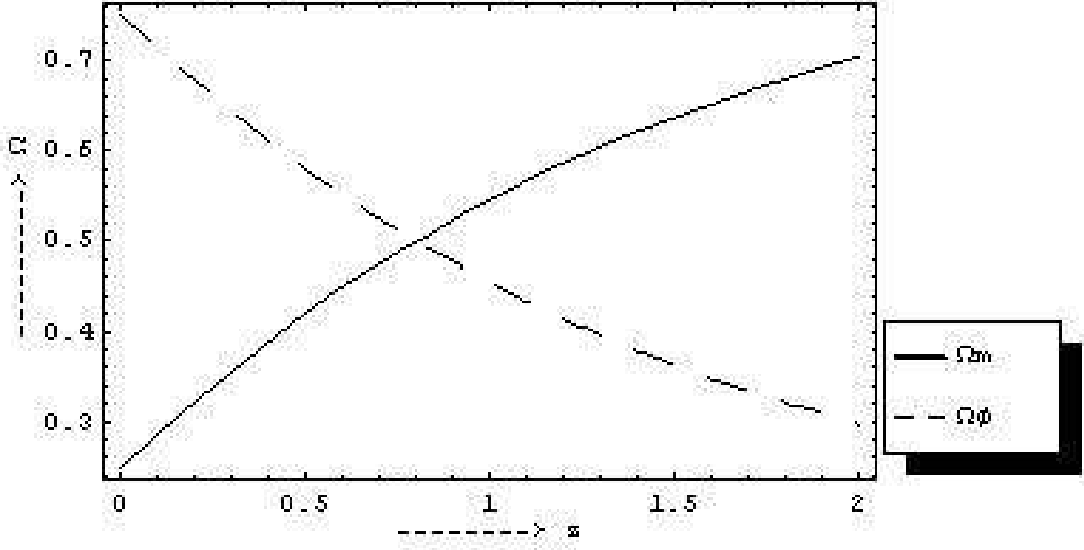,height=3.0in,width=6.0in}}
\caption{Plot of $\Omega$ vs. $z$.}
\label{second_fig}
\end{figure}

 With $ p = -3/2 $, the present model yields
\be
\Omega_{m} = \frac{(1+z){^3}}{[1+(1+z)^{-3/2}]^2} ~~,
\ee

and\\
\be
\Omega_{\phi} = 1 - \Omega_{m}~,
\ee
where $z$ is the redshift parameter given by
\be
1+z = \frac{a_{0}}{a}~,
\ee
 $a_{0}$ being the present value of the scale factor.  $\Omega_{m_{0}}$, the present
value of $\Omega_{m}$, comes out to be 0.25 and $\Omega_{\phi_{0}}$ = 0.75. These values are well within the constraints of 0.2 $\leq \Omega_{m_{0}} \leq$ 0.8  \cite{ref},\cite{vs}. Figure \ref{second_fig} shows that $\Omega_{m}$ increases with $z$, i.e, decreases with the evolution of the universe. $\Omega_{\phi}$ starts dominating over $\Omega_{m}$ roughly at $z$ = 0.8. At the earlier epoch, i.e, at high $z$, $\Omega_{\phi}$ is very small, allowing a conducive matching onto the radiation era for the perfect ambience for nucleosynthesis. \\

\par Unlike Newtonian gravity, general relativity ensures that the pressure also contributes in driving the acceleration of the model. From equations (5) and (6), one has
\be
\frac{\ddot{a}}{a} = -\frac{1}{6} (\rho_{t} + 3 p_{t})
\ee
where $\rho_{t}$ and $p_{t}$ are the total effective density and pressure respectively. If the pressure is connected with the density by the relation
\be
p = w \rho,
\ee
the model will accelerate ( $\ddot{a} > 0$ ) only if
\be
w_{t} = \frac{p_{m} + p_{\phi}}{\rho_{m} + \rho_{\phi}} <  -\frac{1}{3}.
\ee

\begin{figure}[!h]
\mbox{\psfig{figure=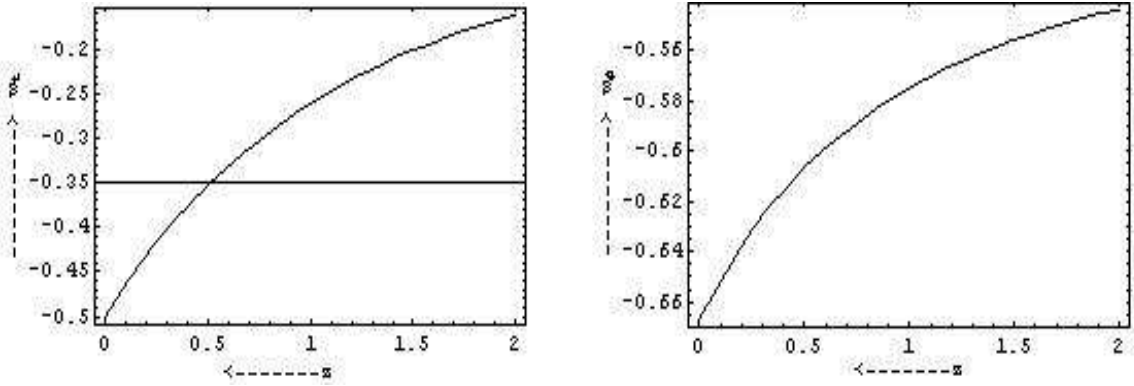,height=2.0in,width=6.0in}}
\caption{Plot of (a) $w_{t}$ vs. $z$  and (b) $w_{\phi}$ vs. $z$.}
\label{third_fig}
\end{figure}

The subscripts `t', `m' and `$\phi$' stand for total, normal fluid distribution
and the Q-matter $\phi$ respectively. For a matter dominated universe, $p_{m} = 0$ and hence $w_{m} = 0$. This particular model gives
\be
w_{\phi} = \frac{p_{\phi}}{\rho_{\phi}} = \frac{1 + (1 + z)^{\frac {3} {2}}}{-1 - 2(1 + z)^{\frac{3}{2}}} 
\ee
\be 
w_{t} = \frac{p_{\phi}}{\rho_{\phi} + \rho_{m}} = -[1 + {(1 + z)^{\frac{3}{2}}}]^{-1}
\ee
The evolution of $w_{t}$ versus $z$  and $w_{\phi}$ versus $z$ are shown in figure \ref{third_fig}, which indicates that $w_{t}$ attains the required value of $-\frac{1}{3}$ or less only close to $z = 0.5$. Beyond that, $w_{t}$ is less negative, and the universe still decelerates. 

\par The value of $w_{\phi0}$, i.e, the value of the equation of state parameter for the scalar field at $z = 0$ as given by the present model and as indicated by figure 3(b) is definitely within the constraint range \cite{ref}.

\begin{figure}[!h]
\mbox{\psfig{figure=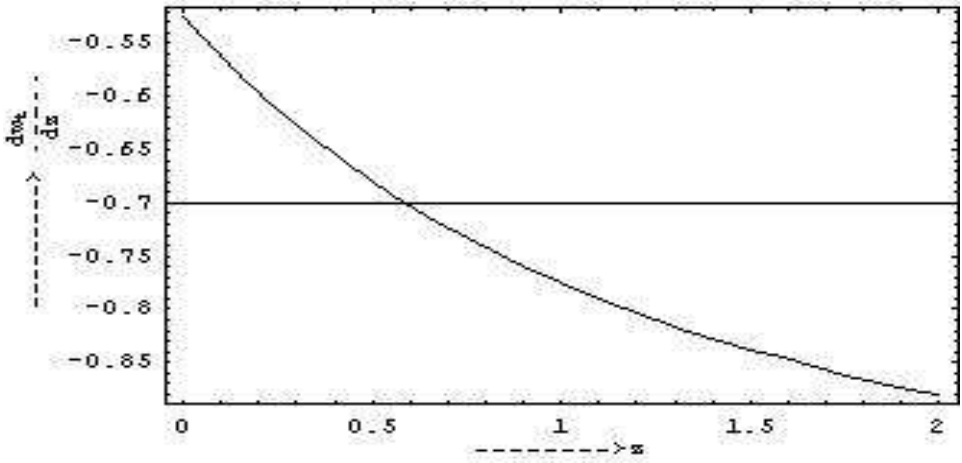,height=2.5in,width=5.0in}}
\caption{Plot of $\frac{dw_{t}}{dz}$ vs. $z$ for spatially flat dust dominated R-W model.}
\label{fourth_fig}
\end{figure}
\par We also plot the rate of change of $w_{t}$ against $z$ ( as shown in figure \ref{fourth_fig} ). It shows that $\frac{dw_{t}}{dz}$ is still negative at the present epoch, but the magnitude of $\frac{dw_{t}}{dz}$ is decreasing.
\par The solution for the scale factor is good enough to allow the density contrast to grow favourably for the formation of large scale structure. Figure (5) shows the growth of linearized density perturbation in this model and evidently indicates that it grows linearly with the scale factor during later stages as expected for the matter dominated epoch \cite{coles}. 

\begin{figure}[!h]
\mbox{\psfig{figure=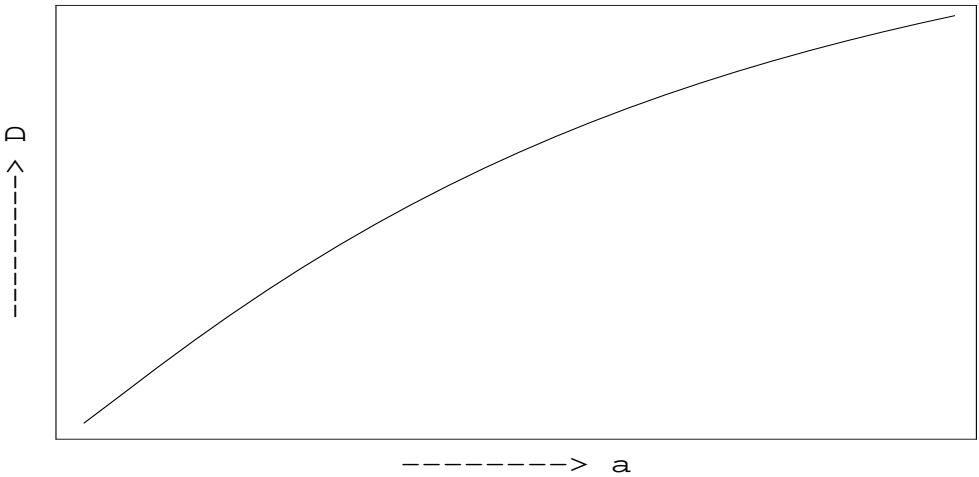,height=2.5in,width=5.3in}}
\caption{Plot of density contrast $D$ vs. $a$ where $ D = \frac{{\rho} - \bar{\rho}}{\bar{\rho}}$.}
\label{fifth_fig}
\end{figure}

\par In the absence of the final form of the quintessence matter, search for 
the relevant form of potential will continue and the present investigation is 
one of them. In view of the high degree of non-linearity of Einstein's 
equations, exact solutions always play a vital role as piecewise solutions 
have the problem of proper matching at different interfaces. The present 
model shows that inspite of the severe constraints imposed by observations, 
one can still find an exact FRW model which gives values for the relevant 
parameters like $q$, $w$, $\Omega_{\phi}$, $\Omega_{m}$ etc. safely within 
the range given by observations. It also has the merit of having a single 
analytical expression for $q = q(a)$, which gracefully transits from its 
positive phase to the negative one and adds to the list of quintessence 
potentials that serve the purpose of modelling a presently accelerating 
universe \cite{v}. Definitely the model has problems, particularly that of 
fine tuning, but infact, all quintessence models have some such problems.

\vskip .2in

\noindent
{\bf Acknowledgement}
 
\vskip .1in

This research is partially supported by DAE (India).

\end{document}